\documentclass[sigconf]{acmart}

\usepackage[T1]{fontenc}
\usepackage{amsfonts}
\usepackage{xspace}
\usepackage{listings}
\usepackage{graphicx}
\usepackage{amsmath}
\usepackage{etoolbox, graphicx, setspace}
\usepackage{tabularx}
\usepackage{subcaption}
\usepackage{enumitem}

\AtBeginDocument{%
  \providecommand\BibTeX{{%
    \normalfont B\kern-0.5em{\scshape i\kern-0.25em b}\kern-0.8em\TeX}}}

\setcopyright{acmcopyright}
\copyrightyear{2022}
\acmYear{2022}

\newcommand{\minihead}[1]{{\vspace{.4em}\noindent\textbf{#1.} }}
\newcommand{\sn}{\textsc{Fixy}\xspace}
\newcommand{\dsl}{\textsc{LOA}\xspace}

\newtoggle{rcolors}
\togglefalse{rcolors}
\newcommand{\colora}[1]{\iftoggle{rcolors}{{\color{blue}{#1}}}{#1}}

\begin{document}

\title{Finding Label and Model Errors in Perception Data With Learned Observation Assertions}

\author{Daniel Kang}
\affiliation{
  \institution{Stanford University}
}

\author{Nikos Arechiga}
\affiliation{
  \institution{Toyota Research Institute}
}

\author{Sudeep Pillai}
\affiliation{
  \institution{Toyota Research Institute}
}

\author{Peter Bailis}
\affiliation{
  \institution{Stanford University}
}

\author{Matei Zaharia}
\affiliation{
  \institution{Stanford University}
}


\begin{abstract}
ML is being deployed in complex, real-world scenarios where errors have
impactful consequences. In these systems, thorough testing of the ML pipelines
is critical. A key component in ML deployment pipelines is the curation of
labeled training data. Common practice in the ML literature assumes that labels
are the ground truth. However, in our experience in a large autonomous vehicle
development center, we have found that vendors can often provide erroneous
labels, which can lead to downstream safety risks in trained models. 

To address these issues, we propose a new abstraction, \emph{learned observation
assertions}, and implement it in a system called \sn. \sn leverages existing
organizational resources, such as existing (possibly noisy) labeled datasets or
previously trained ML models, to learn a probabilistic model for finding errors
in human- or model-generated labels. Given user-provided features and these
existing resources, \sn learns feature distributions that specify likely and unlikely values
(e.g., that a speed of 30mph is likely but 300mph is unlikely). It then uses
these feature distributions to score labels for potential errors. We show that \sn can
automatically rank potential errors in real datasets with up to 2$\times$ higher
precision compared to recent work on model assertions and standard techniques
such as uncertainty sampling.

\end{abstract}



\maketitle

\section{Introduction}

Machine learning (ML) is increasingly being deployed in complex applications
with real-world consequences.  For example, ML models are being deployed to make
predictions over perception data in autonomous vehicles (AVs)
\cite{karpathy2018building}, with potentially fatal consequences for errors,
such as striking pedestrians \cite{wakabayashi2018self}. Thus, quality assurance
and testing of ML pipelines are of paramount concern \cite{zhang2020machine,
amershi2019software, odena2019tensorfuzz, xiang2018verification}.

\begin{figure}[t!]
    \centering
    \includegraphics[width=\columnwidth]{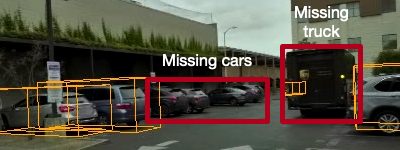}
    \caption{Example of human labels (orange) and missing labels (red) in the Lyft Perception dataset. The black truck highlighted is within 25m of the AV. Such errors can cause downstream issues with perception and planning systems.}
    \label{fig:fig1}
\end{figure}

A critical component of ML deployments is the curation of \emph{high-quality}
training data, in which crowd workers produce labels over data. Similar to how
errors in tabular data results in downstream errors in query results, erroneous
training data (e.g., Figure~\ref{fig:fig1}) can lead to subsequent safety
repercussions for trained models. As such, finding these errors is critical,
which we focus on in this work.

Unfortunately, standard techniques in data cleaning are not well suited for
finding errors in training data. For example, while constraints work well on
tabular data, they are less suited for perception data, e.g., pixels of an
image. As such, we have found it necessary to develop new tools for finding
errors in training data.


Recent work has proposed Model Assertions (MAs) that indicate when errors in ML
model predictions or labels may be occurring \cite{kang2020model}. MAs are
black-box functions over model inputs and outputs that return a quantitative
severity score indicating when an ML model or human-proposed label may have an
error. For example, a MA may assert that a prediction of a box of a car should
not appear and disappear in subsequent frames of a video. MAs can be used to
monitor the ML models in deployment, and to flag problematic data to label and
retrain the model.

However, in our experience deploying MAs in a real-world AV perception training
pipeline, we have found several major challenges. First, users must manually
specify MAs, which can be challenging for complex ML deployments. Second,
calibrating severity scores so that higher severity scores indicate a higher
chance of error is challenging. This is especially important as organizations
have limited resources to evaluate potential errors in ML models or human
labels. Third, ad-hoc methods of specifying severity scores ignores
organizational resources \cite{suri2020leveraging} that are already present:
large amounts of ground-truth labels and existing ML models.

To address these challenges, we propose a probabilistic domain-specific language
(DSL), \emph{Learned Observation Assertions} (\dsl), for specifying assertions,
and methods for data-driven specification of severity scores that leverage
existing resources in ML deployments. We implement \dsl in a prototype system
(\sn), embedded in Python to easily integrate with ML systems.

Our first contribution, \dsl, allows users to specify properties of interest for
perception tasks. \dsl contains three components: data associations, feature
distributions, and application objective functions. \dsl can be used to specify
assertions without ad-hoc code or severity scores by automatically
transforming the specification into a probabilistic graphical model and scoring
data components, producing statistically grounded severity scores.

In our labeling deployment, sensor data across short snippets of time
(\emph{scenes}) are sent to vendors for labeling. These scenes are then audited
for missing labels. These errors are difficult to specify via ad-hoc MAs, so our
DSL supports means of associating observations together: across observation
sources (\emph{observation bundles}, i.e., predictions from different ML models
or sensors) and across time (\emph{tracks}, i.e., predictions of the same object
over time). These associated observations can then be considered jointly when
searching for errors.

Our second contribution is methods of leveraging \emph{organizational
resources}~\cite{suri2020leveraging}, i.e., existing labels and ML models, to
automatically specify severity scores via \dsl. Users specify features over
data, which are used to automatically generate \emph{feature distributions}, and
\emph{application objective functions} (AOFs) to guide the search for errors.
Feature distributions take sets of observations and output a probability of
seeing a feature of the input. For example, a feature distribution might take a
3D bounding box of a car and return the likelihood of encountering that box
volume. AOFs transform feature distribution values for the application at hand.
For example, if we wish to find likely tracks (e.g., a track missed by human
labels), the AOF could simply return the feature distributions' value. If we
wish to find unlikely tracks (e.g., a ``ghost'' track that an ML model
erroneously predicts), the AOF could return one minus the feature distributions'
value. 


We evaluate \sn on two real-world AV datasets, the publicly available Lyft Level
5 perception dataset \cite{lyft2019} as well as an internally collected dataset.
Both datasets were annotated by leading commercial labeling vendors. Despite
best efforts from these vendors \cite{cheng2019training}, we find a number of
labeling errors via \sn, some of which could cause safety violations (e.g., in
Figures~\ref{fig:fig1} and \ref{fig:lyft_errors}). We first show that \sn can
achieve recalls of up to 75\% when searching for errors in these datasets, while
achieving 2$\times$ higher precision for finding label error than hand-crafted
MAs. Furthermore, \sn can find novel errors in ML models that the hand-crafted
MAs in previous work are unable to find, and finds high-confidence errors that
uncertainty sampling misses.

In summary, our contributions are
\begin{enumerate}[itemsep=0em, parsep=0em, topsep=0em, leftmargin=1.5em]
    \item \dsl, a probabilistic DSL with syntax and semantics for validating
    observations over complex perception data.

    \item Methods for leveraging organizational resources (in the form of
    existing ML models and labels) to automatically tune feature distributions
    and detect errors.

    \item An empirical evaluation of our implementation \sn, showing it can
    outperform baselines for detecting errors even in commercially generated and
    vetted label data.
\end{enumerate}

\begin{figure*}[t!]
    \begin{subfigure}{\textwidth}
    \includegraphics[width=\textwidth]{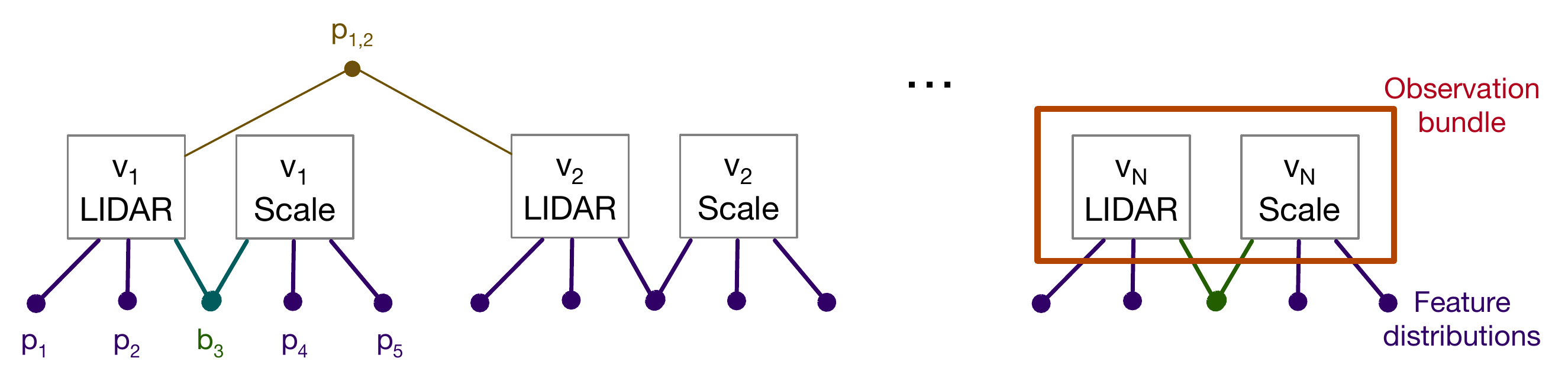}

    \caption{Schematic of a track that contains observations from LIDAR ML model
    predictions and from human-proposed labels. We show examples of feature
    distributions for observations ($p_1$,  $p_2$, $p_4$, and $p_5$), bundles
    ($b_3$), and transitions ($p_{1,2}$).}

    \label{fig:track-schematic}
    \end{subfigure}

    \centering
    \begin{subfigure}{0.24\textwidth}
        \includegraphics[width=\textwidth]{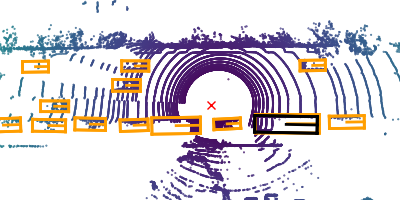}
        \caption{$t=1$ ($v_1$ model and $v_1$ human)}
    \end{subfigure}
    \begin{subfigure}{0.24\textwidth}
        \includegraphics[width=\textwidth]{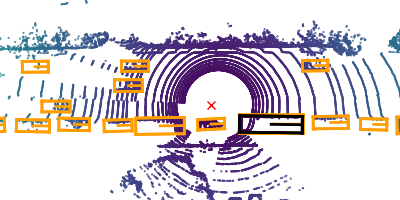}
        \caption{$t=2$ ($v_2$ model and $v_2$ human)}
    \end{subfigure}
    \begin{subfigure}{0.24\textwidth}
        \includegraphics[width=\textwidth]{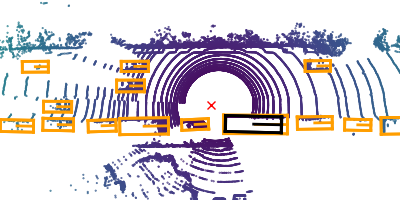}
        \caption{$t=3$ ($v_3$ model and $v_3$ human)}
    \end{subfigure}
    \begin{subfigure}{0.24\textwidth}
        \includegraphics[width=\textwidth]{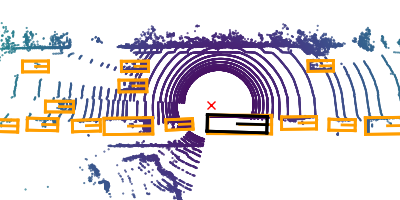}
        \caption{$t=4$ ($v_4$ model and $v_4$ human)}
    \end{subfigure}
    \caption{Example of the factor graph (top) and corresponding LIDAR point
    cloud data (bottom). \colora{We show here and throughout the birds eye view of the
    LIDAR frame. Concentric circles represent equidistance from the sensor.
    Reflected points from objects are shown as dots.} The track is in black and other human-proposed labels are in orange for reference.}
    \label{fig:track_example}
\end{figure*}

\section{Example and Background}
\label{sec:background}

\minihead{ML workflow}
As an example, we describe the ML deployment pipeline for our AVs, focusing on
labeling data for perception systems. Other organizations deploy similar
pipelines, e.g., as documented by \citet{karpathy2018building}.

Our AV deployment pipeline is a continuous process, in which ML models are trained, tested, and deployed on vehicles. Because ML models are continuously exposed to new and different scenarios, we continuously collect and label data, which is subsequently used to develop and retrain ML models \cite{baylor2017tfx}.

Label quality is of paramount concern: erroneous labels can lead to downstream
errors, which in turn can lead to safety violations. Vendors that provide labels
are not always accurate, which is in contrast to the large body of work that
assumes datasets are ``gold.'' For our perception system, the most egregious
errors are when objects are entirely missed in labeling. \colora{We show examples of
missing labels in Figure~\ref{fig:fig1}, in which a truck and several cars were
missed by the human labeler.}

To address label quality issues, our organization has expert auditors who audit the vendor-provided labels. Unfortunately, it is too expensive to audit every data point, so we have developed \sn, which enables ranking datapoints that are likely to be erroneous and allows better utilization of auditing resources.

\minihead{Model assertions}
MAs are user-provided, black box functions over ML model inputs and outputs that
indicate if the ML model has an error \cite{kang2020model}. MAs can be deployed
at test time to indicate possible errors so corrective actions can be taken.
They can additionally be used to select data that produces errors for labeling,
e.g., as studied by \citet{kang2020model}, as many organizations continuously
collect data to label.

Unfortunately, MAs are specified in an ad-hoc manner. They require users to
write programs to specify exactly what forms of errors occur and ad-hoc severity
scores to indicate the likelihood of an error. We have found that these ad-hoc
procedures can be challenging to use.



\minihead{Factor graphs}
\sn generates graphical models from data and feature distributions. We
specifically consider factor graphs due to the ease of representing data and
distributions \cite{kschischang2001factor}.

Given a set of random variables $X = \{X_1, \dots, X_n\}$, a factor graph represents a factorization of a joint distribution $g(X_1, \dots, X_n)$. Assume that the joint distribution can be factorized in terms of a set of functions $f_j$, which we will call \emph{factors}, and $S_j \subseteq X$
\begin{align}
g(X_1, ..., X_n) = \prod_{j = 1}^{m} f_j(S_j).
\end{align}

Formally, we can represent a factor graph as a graph $G = (X, F, E)$, where $X$ and $F$ are two disjoint sets of nodes. The graph is \emph{bipartite}, meaning that each edge connects a node in $X$ to a node in $F$, but no edge connects nodes in $X$ among themselves nor nodes in $F$ among themselves.
For every factor $f_j \in F$, there is an edge that connects it to $X_i$ if and only if $X_i \in S_j$ in the factorization of $g$.

We consider specific factor graphs that are automatically generated by \sn, as
described later in this paper.

\minihead{LIDAR}
We extensively use and show LIDAR data and predictions over LIDAR data as
examples of missing human labels or ML model predictions. \colora{LIDAR is generated by
pulsing light and timing the returns of the pulsed
light~\cite{wandinger2005introduction}. With accurate timings, LIDAR data gives
accurate distance measurements and are represented as point clouds. In this
paper, we show birds-eye view of LIDAR data: concentric circles indicate same
distances from the LIDAR sensor and we draw predicted boxes over the scenes. We
show an example in Figure~\ref{fig:track_example}. LIDAR figures in white
background are from the Lyft Level 5 perception dataset~\cite{lyft2019} and
figures in black background are from our internal dataset.}

\section{System Overview}
\label{sec:overview}

\colora{
\minihead{Goals}
\sn aims to enable users to find errors in ML labeling pipelines and in ML
models, primarily in the form of missing labels. In particular, \sn aims to
reduce manual effort by only requiring users to specify natural quantities
(e.g., box volume, velocity) as opposed to specifying the exact form of errors
as model assertions expect users to do.

}

\minihead{Inputs and outputs}
We first denote human-proposed labels and ML model outputs as ``observations.''
As input, \sn takes a set of observations. As output, \sn returns a ranked list
of (potentially a subset of) observations, where higher ranked observations are
ideally more likely to contain errors.

Offline, \sn takes already-present labels to learn feature distributions over
features of the observations. \sn will then use this data to rank potential
errors.

\begin{figure}[t!]
  \includegraphics[width=\columnwidth]{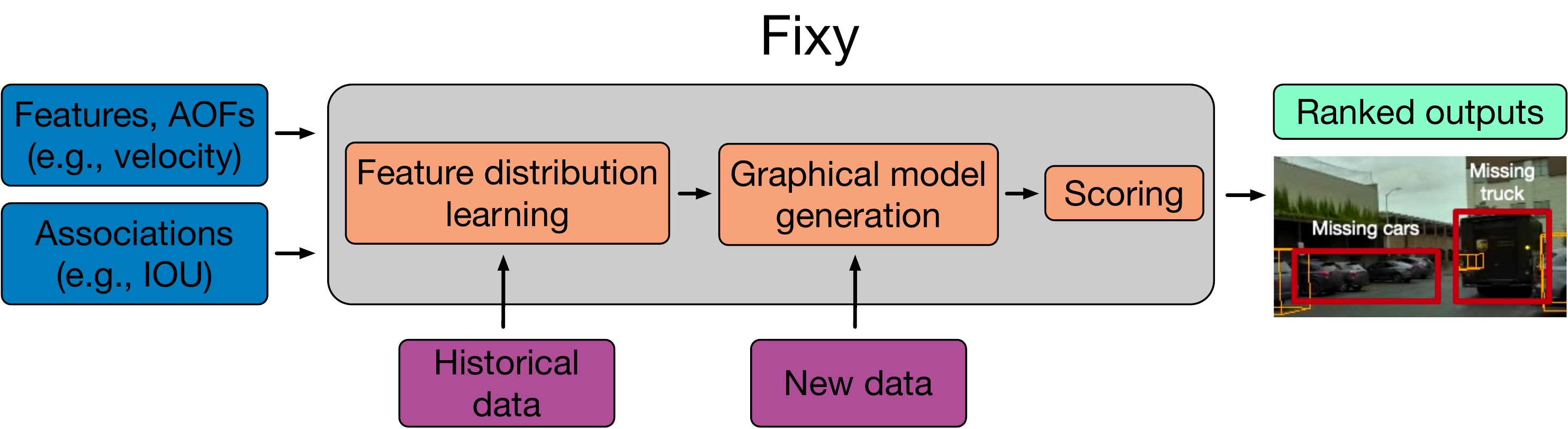}

  \caption{System diagram for \sn. Users provide features over perception data
  (e.g., box volume) and associations between observations.  Given these inputs,
  \sn will learn feature distributions, generate graphical models, score new
  data, and output potential errors.}

  \label{fig:sys-diagram}
\end{figure}

\minihead{\sn components}
\sn consists of: a DSL for specifying relations between observations and feature
distributions, a component to learn feature distributions, a scoring component,
and a runtime engine. \sn's DSL allows users to specify how feature
distributions and observations interact. Its distribution learning component fits
distributions over existing observations. Its scoring component scores
observations or groups of observations by likelihood. Finally, its runtime
engine ranks observations or groups of observations.

We show a system diagram in Figure~\ref{fig:sys-diagram}. Users need
only provide the features (and data to be ranked). Once the feature
distributions are learned, \sn will rank potential errors for auditing.

\minihead{Workflow}
\sn contains an offline (distribution learning) and online (error ranking)
phase. In the offline phase, \sn will take existing organizational resources in
the form of existing labels to learn feature distributions. In the online phase,
\sn will rank potential errors.

We have found that users of label checking tools are often non-experts in coding
and ML tools, so we have opted for simplicity in \dsl. Thus, a user of \sn need
only specify features and optionally AOFs. In particular, many features are
\emph{already computed} for use in other pipelines so can be reused (e.g.,
object volume, velocity, and distance from vehicle). Thus, the features can be
specified in few lines of code, as we show below.

\minihead{Worked example}
Consider the use case of finding missing human labels of 3D bounding boxes over
LIDAR point cloud data. For example, Figure~\ref{fig:fig1} show several missing
cars and a missing truck.

In this example, we have two sources of observations: predictions from an ML model
and human labels. To find these errors, the user will: 1) associate observations
and 2) specify features. Then, \sn will automatically score and rank potential
errors.

The analyst first associates observations within a time step (i.e., overlapping
model predictions and human labels) and between adjacent timesteps (i.e.,
objects across time). To do so, the user can specify that observations with high
box overlap are associated. While this is provided by default by \sn, the user
can also write a short amount of code using the intersection over union (IOU):
\begin{lstlisting}
class TrackBundler(Bundler):
  def is_associated(self, box1, box2):
    return compute_iou(box1, box2) > 0.5
\end{lstlisting}

The analyst then specifies features. As a concrete example, the analyst may
specify a feature that computes box volume. The user need only provide code to
compute the box volume: \sn will learn distribution of box volumes and use it to
find anomalous boxes.
\begin{lstlisting}
# KDEObsDistribution takes features and learns a
# KDE density estimator over the features
class VolumeDistribution(KDEObsDistribution):
  def feature(self, box):
    vol = box.width * box.height * box.length
    return vol
\end{lstlisting}
The user can also specify other features, such as object velocity. The two code
snippets above (and another other features the user wishes to specify) are all
that a user need to provide to \sn.

Given the associations between observations and the features, \sn will learn the
likelihood of encountering specific feature values offline, using
already-collected resources.

Once  these feature distributions are learned, \sn will score and rank new data,
ideally with potential errors ranked higher. Concretely, consider
Figure~\ref{fig:fig1}.  Although not shown, an ML model highlighted the truck in
a time-consistent way.  Since the track is highly consistent, \sn returns a high
likelihood of an error.  An expert auditor can then verify if the potential
error is actually a missing label.

\section{Learned Observation Assertions}
\label{sec:dsl}

The \dsl DSL provides a simple means of specifying associations between
observations and specifying associations between observations and feature
distributions. Intuitively, applications that contain observations over time and
over multiple modalities/models may have observations that are associated across
time/modalities. Furthermore, feature distributions may operate over individual
observations or groups of observations. We show an example of a compiled \dsl
graph and corresponding sensor data observations in
Figure~\ref{fig:track_example}.

In this section, we provide a formal description of \dsl. However, users
interface with \dsl via a Python library. In particular, users only need to
specify features over which distributions are learned and methods of associating
observations. Our implementation provides class interfaces where users can
override the feature computation (for the feature distributions) and the
association method (for associating observations). We show an example in
Section~\ref{sec:overview} of the code the user needs to provide.

\subsection{Overview}
\dsl contains elements for allowing users to specify how observations interact
with each other and how feature distributions interact with observations. Our
implementation of \dsl is embedded in Python for ease of integration with
standard ML packages.  Since perception data often contains spatial and temporal
components, we allow users to construct \emph{observation bundles} within a
single time step and \emph{tracks} across time. We collectively refer to
observations, bundles, and tracks as OBTs. \dsl then allows features to be
specified over any OBT. Finally, the user can specify application objective
functions (AOFs) over any feature distribution.

\subsection{Scene Syntax}

\minihead{Overview}
We consider \emph{scenes} of data, which consists of \emph{observations} and
\emph{features} over these observations. Our syntax consists of specifying how
observations relate to each other within a scene and how features relate to groups
of observations.


\minihead{Formalism}
A scene consists of a set of tracks. Each track contains a set of observation
bundles. An observation bundle contains observations from different modalities,
such as LIDAR, vision, and for offline data, human proposals of labels. We
summarize our syntax notation in Table~\ref{tab:syntax}.

Formally, we denote the scene (i.e., set of tracks) as $s = \{ \tau \}$.  Each
track consists of an indexed sequence of observation bundles, $\tau = (\beta_0,
\dots, \beta_n)$. Each observation bundle consists of a set of observations,
$\beta = \{\omega_{\tau,\beta}\}$.

In order to reason about erroneous or unusual artifacts in the perception
system, we define features over the elements of the scene. Users can assign
features to any of the elements of the scene; these assignments are often done
automatically (e.g., a volume feature would apply over every observation).
Concretely, features can be over observations, observation bundles, tracks, or
entire scenes. 

Formally, $\pi$, the feature function, maps each element to its features. For
example, $\pi(\omega_{\tau,\beta})$ are the features assigned to the observation
in track $\tau$ in bundle $\beta$, which could be a feature on the volume of the
object detected. Similarly, $\pi(\tau)$ assigns track $\tau$ its features, which
could be the total number of observations within that track.

In addition to features over discrete groups of observations, we provide syntax
for features over adjacent observations within a track (``transition features''),
i.e., $\pi(\beta_i, \beta_{i+1})$. As a concrete example, we have implemented a
transition feature for the estimated instantaneous velocity. We note that this
syntax is for convenience, as it could be implemented via track features.
Nonetheless, we have found it useful in our applications to allow for transition
features.

Finally, AOFs can be specified over any feature distribution. These AOFs are
numeric transformations of the returned feature distribution score, e.g., the
identity function, the zero function, or $f(x) = 1 - x$.


%
%

\subsection{Generating Graphical Models}

\sn will compile the scene, feature distributions, and AOFs to a graphical
model, which is used to score groups of observations. \sn uses these scores to
flag potential errors. 

To compile a scene, \sn will create nodes for each observation and feature
distribution. Then, \sn will create edges between each feature distribution and
the observation it applies over. If a feature distribution applies to a single
observation, \sn will create a single edge. If a feature distribution applies to
a group of observations (e.g., an observation bundle or track), \sn will create
one edge between each observation in the group and the feature distribution.

Once the graphical model is constructed, \sn can then score any OBT. \sn will
score an observation by the negative log-likelihood implied by its feature
distributions. The score of a group of observations is the sum of the scores of
the observations, normalized by the number of feature distributions. We defer the full
discussion of scoring and a worked example to Section~\ref{sec:scoring}.

\begin{table}[t!]
    \centering
    \begin{tabular}{c|l}
        Syntactic element & Meaning  \\ \hline
        $s$      & Scene \\
        $\tau$   & Track \\
        $\beta$  & Observation bundle \\
        $\omega$ & Observation \\
        $\pi$    & Feature mapping function
    \end{tabular}
    \caption{Table of syntactic elements in \sn's DSL.}
    \label{tab:syntax}
\end{table}

\section{Feature Distributions}

A key component to scoring OBTs are the feature distributions. Both our AV
deployment and other organizations deploying ML collect large amounts of
training data. This training data contains labels (potentially with errors),
which can be used to fit empirical distributions to the features. We leverage
these existing labels in this work, as they come at no additional cost.

To fit these feature distributions, \sn takes as input scalar or vector valued
features over OBTs. For example, a feature over an observation may take a
bounding box and return the volume of the box \colora{as described in
Section~\ref{sec:overview}}. The user may also manually specify feature
distributions to rank severity (e.g., distance of an object to the AV) or to
filter certain instances (e.g., only search for errors in detecting
pedestrians).
Finally, \sn takes an optional AOF, which can be applied per feature or over the
resulting score.

We describe the feature types, their specification, and how \sn fits them below.

\begin{figure*}[t!]
    \centering
    \includegraphics[width=0.32\textwidth]{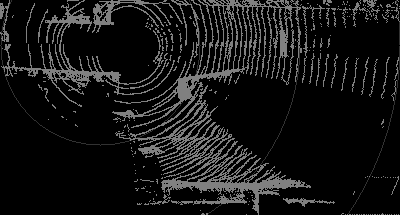}
    \includegraphics[width=0.32\textwidth]{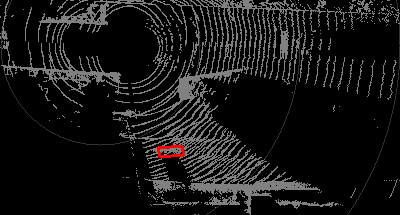}
    \includegraphics[width=0.32\textwidth]{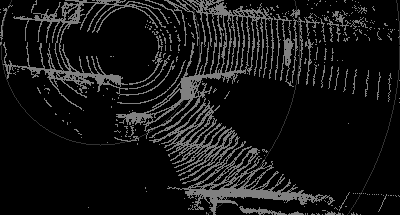}
    \includegraphics[width=0.32\textwidth]{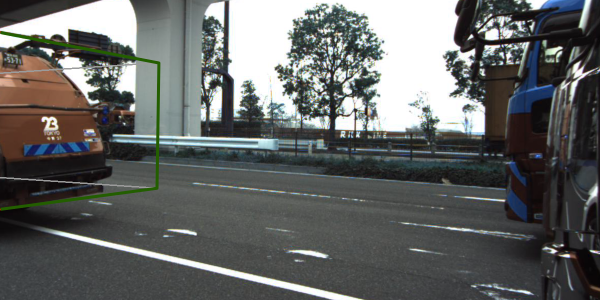}
    \includegraphics[width=0.32\textwidth]{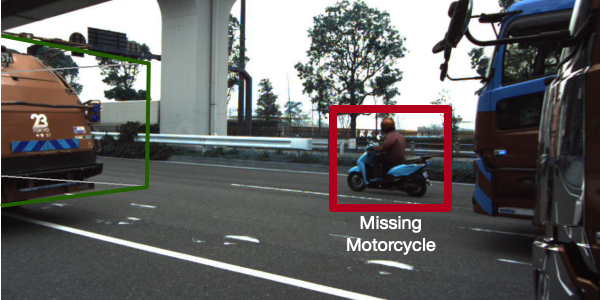}
    \includegraphics[width=0.32\textwidth]{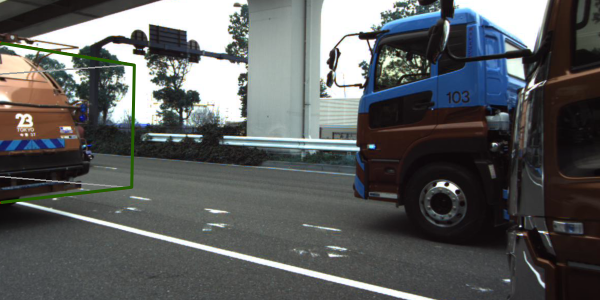}
    \caption{Example of a motorcycle (highlighted in red) missed by human proposals. We show both the LIDAR point cloud data (top) and the camera view (bottom). The motorcycle is occluded by other vehicles, so only appears for a short period of time (<1 second). Nonetheless, it is close to the AV and is thus important to predict.}
    \label{fig:motorcycle}
\end{figure*}

\subsection{Feature Types}
\sn contains features over OBTs and transitions. While transition features can be
implemented as track features, we provide a syntactic element for ease of use. 

\sn's first feature type are features over single observations. Each feature is
associated with a specific observation type (e.g., a feature over a LIDAR model
prediction). These features are typically used to specify time-independent
information over the predictions. For example, a feature may take a 3D box
prediction from a LIDAR model and return the box volume. The observation feature
would be over box volumes in this case.

\sn's second feature type are features over observation bundles. These features are
typically used to specify consistency between observations of the same object in
a single time step. For example, consider the intuition that observations within
bundles should agree on object class. To specify this, a user could provide a
feature that returns 0 if all the classes agree and 1 otherwise. The feature would
then learn the Bernoulli probability of the class agreement between observation
types.

\sn's third feature type are features between observations or bundles in adjacent
time steps within a track. These features are typically used to specify
information over time-dependent quantities or consistency. For example, a
feature could specify the velocity estimated by box center offset.

\sn's fourth feature type are features over entire tracks. Although rare, these
features can be used to normalize scores over entire tracks.

\subsection{Learning Feature Distributions}

Given features, \sn can automatically fit feature distributions over existing
training datasets. To fit feature distributions, \sn takes a function that
accepts a list of scalars/vectors and returns a fitted distribution. By default,
\sn uses a kernel density estimator (KDE) to learn feature distributions over
the features. In some cases, other types of distributions are appropriate (e.g.,
discrete distributions): the user can override our default KDE estimator in
these cases.

To learn feature distributions given a set of scenes, \sn first exhaustively
generates the features over the data and collects the scalar or vector values.
Then, for each feature, \sn executes the fitting function over the scalar/vector
values.

\colora{
We note the density estimators have hyperparameters. We have found that default
hyperparameters work in all cases we tried, so we defer exploring
hyperparameters to future work.

}

\subsection{Application Objective Functions}
AOFs wrap data feature distributions to transform them into an
application-specific probabilities to guide the search for labeling errors. As
such, they take scalar values and return scalar values. The most common
operations are taking the inverse and setting the probability to 0/1 under
certain conditions.
For example, when searching for likely tracks, the application objective
function may be the identity. In contrast, when searching for unlikely tracks,
the application objective function may invert the probability.

\section{Scoring Relative Plausibility}
\label{sec:scoring}

Given the compiled factor graph, \sn can score any OBT.
\sn will first score the observations via the sum of log likelihood of the
feature distributions transformed by the application objective functions.
Formally, given the AOFs $f_i$, the score for an observation $\omega$ is
\begin{align}
\sum_{\pi_i \in \pi(\omega)} \ln \left( f_i(\pi_i(\omega)) \right).
\end{align}

The score of any component in the graph is the sum of the scores of the
observations, normalized by the total number of features that connect to the
component. We normalize by the number of features so that components of
different sizes are comparable (e.g., a track with 10 observations compared to a
track with 100 observations).

Consider the missing truck in Figure~\ref{fig:fig1} and suppose the ML model
predicted it in two adjacent time steps ($t=1,2$) for simplicity. Suppose the
predicted box volumes are 44.8 m$^3$ and 45.9 m$^3$, and the predicted
velocity was 2 m/s. In this case, the scores of the volumes may be 0.37 and 0.39
respectively, and the score for the velocity may be 0.21. The score for the
track would be $(\ln(0.37) + \ln(0.39) + \ln(0.21)) / 3 = -1.17$. Since \sn only uses
the score to rank, this would be compared to other scores.

\section{Applications}
\label{sec:applications}

We provide examples of applying \sn to finding different kinds of errors in ML
applications. For all applications, we assume that the predictions are 3D
bounding boxes over LIDAR point cloud data. We further assume access to two
features: an observation feature over box volume and a transition feature over
estimated velocity.

\minihead{Finding missing tracks}
In this application, we are interested in finding tracks that human proposals
missed entirely. For example, Figure~\ref{fig:motorcycle} shows a motorcycle
close to the AV but is only visible for a short period of time due to occlusion.
It is important to find such errors as this may cause ML models to misclassify
motorcycles at deployment time.

To find such errors, we additionally execute a 3D bounding box prediction model over the data. Given the ML model predictions, we associate ML model predictions and human proposal in the same frame if they have high box overlap.

\begin{figure}[t!]
    \centering
    \includegraphics[width=0.49\columnwidth]{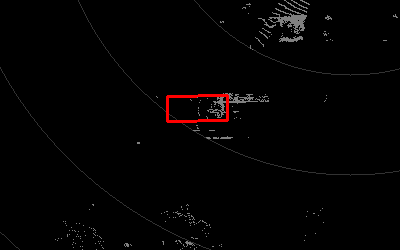}
    \includegraphics[width=0.49\columnwidth]{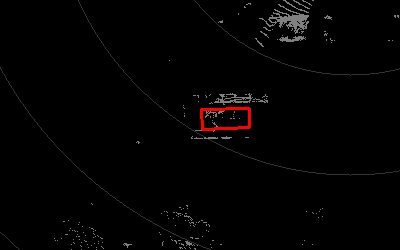}
    \caption{Example of an unlikely track. Predictions are inconsistent within a track, suggesting that they are spurious.}
    \label{fig:bad_track}
\end{figure}

The AOF zeros out any track that contains any human proposals. The remaining
tracks contain only model predictions and are scored as usual, with the
intuition that consistent predictions from the model are likely to be correct.
We show an example of a high probability track (Figure~\ref{fig:motorcycle}) and
low probability track (Figure~\ref{fig:bad_track}).

\begin{figure}[t!]
    \centering
    \includegraphics[width=0.49\columnwidth]{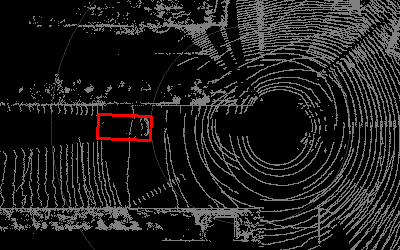}
    \includegraphics[width=0.49\columnwidth]{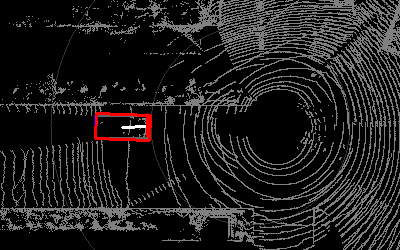}
    \caption{Example of missing human label within a track that \sn can find.
    The left panel only contains an ML model prediction while the right contains
    both a human label and an ML model prediction.}
    \label{fig:missing_obs}
\end{figure}

\begin{figure}[t!]
    \centering
    \includegraphics[width=0.49\columnwidth]{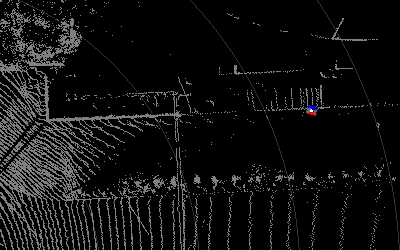}
    \includegraphics[width=0.49\columnwidth]{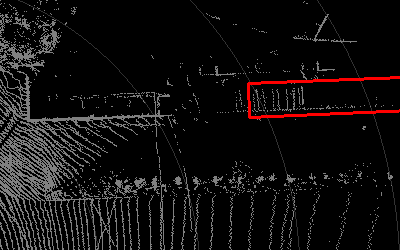}
    \caption{Example of a low probability bundle that would have a low rank. The
    box of the person and truck highly overlap, but are strongly inconsistent in
    box volume.}
    \label{fig:mis_specification}
\end{figure}

\minihead{Finding missing labels within tracks}
We are interested in finding errors in labels proposed by humans that should
belong to an existing track. For example, Figure~\ref{fig:missing_obs} shows a
car trailing the AV, where the first frame is missing the car box.

To find such errors, we use the 3D bounding box prediction model's predictions.
The association of observations into bundles is done as above. The AOF zeros out
the probability of any bundle that contains a human proposal and any track that
does not contain any human proposals. Thus, the remaining bundles only contain
ML model predictions and are in tracks that contain at least one human proposal.

The remaining bundles are scored as usual, with the intuition that predicted
boxes that produce high probability bundles are likely to be correct
predictions. We show an example of a high probability bundle
(Figure~\ref{fig:missing_obs}) compared to a low probability bundle
(Figure~\ref{fig:mis_specification}).

\minihead{Finding erroneous ML model predictions}
In this application, we are interested in finding erroneous ML model
predictions. As such, we assume there are no human proposals. We show an example
of an erroneous track in Figure~\ref{fig:model_error}, where the truck is
inconsistently predicted.

The AOF inverts the probability of each feature, with the goal of inverting the
ranking of the tracks that are likely to be correct and the tracks that are
likely to be incorrect. 

\section{Evaluation}

We evaluated \sn on whether it can find errors in ML pipelines. We show that \sn
can find errors in human-proposed labels that are difficult to specify with
ad-hoc MAs and novel errors in ML model predictions that prior work in ad-hoc
MAs and uncertainty sampling cannot find.

\subsection{Experimental Setup}
\minihead{Datasets}
We evaluated \sn on two AV perception datasets: an internal dataset from our
research organization \colora{and the publicly available Lyft Level 5 perception
dataset \cite{lyft2019}. The Lyft dataset has been used to develop models
\cite{zhu2019class} and host competitions \cite{shet2019lyft}.} Both datasets consists of many
scenes of LIDAR and camera data that were densely labeled with 3D bounding boxes
by leading external vendors for human labels (``human-proposed labels'').

\colora{
We executed \sn on 46 scenes from the Lyft dataset (the entire validation set,
i.e., not seen at training time) and 13 scenes from our internal dataset.
Additionally, we asses the recall of \sn on a scene from our internal dataset
that we vetted carefully.

}

The class labels, sampling rate, and physical sensor layout differ
between the two datasets, showing that \sn can apply across dataset
characteristics.

\minihead{Observation sources}
We used three sources of observations over the data: human-proposed labels,
LIDAR ML model predictions \cite{lang2019pointpillars, zhu2019class}, and expert
auditor labels. All sources predict 3D bounding boxes. We focus on the common
classes of car, truck, pedestrian, and motorcycle.


\minihead{Features}
We used the features shown in Table~\ref{table:features}. These features consist of
features that were automatically learned from data (volume, velocity, count) in
addition to features for selecting more egregious errors (distance, model only).

\minihead{Baselines}
We compared against manually designed ad-hoc MAs developed by
\citet{kang2020model} and uncertainty sampling. The ad-hoc MAs were designed to
find errors in similar settings to ours, across both human-proposed labels and
ML model predictions. Uncertainty sampling is commonly used in active learning
\cite{settles2009active}.

As we describe (Section~\ref{sec:overview}), the users of \sn are typically
non-experts in coding and ML tools. As such, we focus on simple ad-hoc MAs
(e.g., ones developed by \citet{kang2020model}) and low-code features in \sn.
Each feature required fewer than 6 lines of code to implement.

Furthermore, both datasets were vetted by leading vendors for human labels.
Thus, we find errors that were not found in an external audit.

\colora{
\minihead{Runtime}
Since \sn is primarily designed to operate in batch on ingested data and shown
to auditors, the latency per scene is not a critical metric. \sn executes in
under five seconds on a single CPU core for processing a 15 second scene of
data. For context, auditing a scene of data takes orders of magnitude longer, so
this latency can easily be hidden.

}

\begin{table}[t!]
    \centering
    \begin{tabularx}{\columnwidth}{llX}
        Name     & Type & Description \\ \hline
        Volume   & Obs.   & Class-conditional box volume \\
        Distance & Obs.   & Distance to AV \\
        Model only & Bundle & Selects bundles with model predictions only \\
        Velocity & Trans. & Class-conditional object velocity \\
        Count    & Track  & Filters tracks with two or fewer obs.
    \end{tabularx}
    \caption{Description of features we used in this evaluation. Model only and
    count were manually specified features.}
    \label{table:features}
\end{table}

\subsection{\sn can Find Missing Tracks}
We investigated whether \sn could find errors in vendor-proposed labels. We
searched for tracks that were entirely missed by human proposals, as these
errors are the most egregious.

\minihead{Experimental setup}
To find tracks entirely missed by human labelers, we associated LIDAR
observations and human observations by box overlap within the same frame and
associated observations within a track by box overlap across time. We further
deployed the features described above.
\colora{For the ad-hoc MA baseline, we used the ``consistency''
assertion as described by \citet{kang2020model}.} For comparison purposes, we
ordered the ML model predictions randomly and by model confidence.

\begin{table*}[t!]
    \centering
    \begin{tabular}{lllll}
        Method           & Dataset  & Precision at top 10 & Precision at top 5 & Precision at top 1   \\ \hline \hline
        \sn              & Lyft     & \textbf{69\%} & \textbf{70\%}  & \textbf{67\%} \\
        Ad-hoc MA (rand) & Lyft     & 32\%          & 30\%           & 24\% \\
        Ad-hoc MA (conf) & Lyft     & 39\%          & 40\%           & 39\% \\ \hline
        \sn              & Internal & \textbf{76\%} & \textbf{100\%} & \textbf{100\%} \\
        Ad-hoc MA (rand) & Internal & 49\%          & 64\%          & 66\% \\
        Ad-hoc MA (conf) & Internal & 71\%          & 86\%          & 66\% \\
    \end{tabular}
    \caption{Precision at top 10 of \sn and ad-hoc MA baselines for finding
    tracks missed by humans. \sn used features and ad-hoc MAs used the consistency
    assertion described by \citet{kang2020model}. \sn outperforms baselines by
    up to 2$\times$.}
    \label{table:missing-tracks}
    \vspace{-1em}
\end{table*}

We manually checked the top 10 potential errors as proposed by \sn and ad-hoc
model assertions (in some cases, fewer than 10 potential errors were flagged; we
use the maximum number in these cases). We measured the precision among these
potential errors, where a higher precision indicates that there are more errors
within the top 10 proposals. For the Lyft dataset, we measured the precision
across every scene in the validation set (i.e., data that was not seen during
model training) that we discovered errors. For our internal dataset, we focused
on the scene that failed audit.

\minihead{Results: recall}
To assess the recall of \sn, we exhaustively audited a 15 second scene from our
internal dataset. It contained 24 missing tracks. In this scene, \sn achieved a
recall of 75\%, finding 18 of the missing tracks in the top 10 ranked errors
per-class. We believe this result is reflective of the larger dataset.

We further manually searched for errors in the Lyft dataset and found errors in
32 of the 46 scenes. Unfortunately, due to the sheer number of errors in the
Lyft dataset, we were unable to perform recall experiments on the level of
boxes. However, LOA found errors in \emph{100\% of the scenes with errors} in
the top 10 ranked errors.

\minihead{Results: precision}
\sn outperforms on finding errors on precision in both datasets
(Table~\ref{table:missing-tracks}) by aggregating information across
observations in tracks, which is difficult to do with ad-hoc MAs.

\begin{figure*}[t!]
    \centering
    \begin{subfigure}{0.33\textwidth}
        \includegraphics[width=\textwidth]{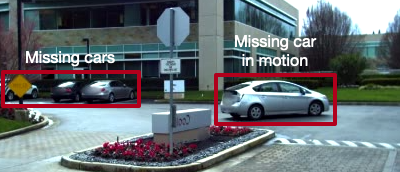}
        \caption{Example of a missing car in motion.}
    \end{subfigure}
    \begin{subfigure}{0.33\textwidth}
        \includegraphics[width=\textwidth]{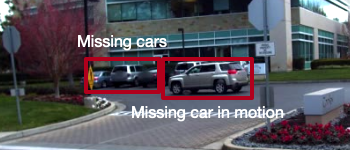}
        \caption{Example of a missing car in motion.}
    \end{subfigure}
    \begin{subfigure}{0.33\textwidth}
        \includegraphics[width=\textwidth]{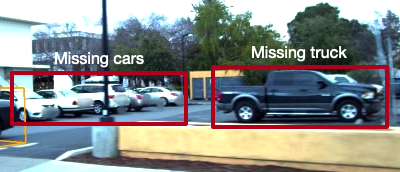}
        \caption{Example of missing cars.}
    \end{subfigure}
    \caption{Examples of labeling errors in the Lyft dataset. The missing objects in these examples can be within 20 meters the autonomous vehicle and several are in motion: vehicles in motion are the most important to detect.}
    \label{fig:lyft_errors}
\end{figure*}

We show three examples of errors \sn found in the Lyft dataset in
Figure~\ref{fig:lyft_errors}. Many of these errors are close to the AV and are
clearly visible. These errors are problematic because they can confuse ML models
and could potentially cause downstream issues.

\minihead{Discussion}
To further contextualize our results, we note that \sn uncovered an error that
was missed by an internal audit. Specifically, the motorcycle track described in
Section~\ref{sec:applications} (Figure~\ref{fig:motorcycle}) was not found in
our initial internal audit. Given the short time period the motorcycle was
visible, it can be difficult to find for both crowd workers and auditors.
\colora{Nonetheless, it is critical to be accurately labeled for two key
reasons. First, clean training data is critical for liability purposes should an
accident occur. Second, the motorcycle is close to the autonomous vehicle, which is
especially problematic for downstream planning.}

We note that our internal model does better than the public model. This is
primarily because the Lyft dataset is very noisy: our internal model was trained
on \emph{already audited} data, which is of higher quality and results in more
calibrated model predictions. These results highlight the need for high quality
data: noisy data results in lower performing models.

Furthermore, the open-sourced Lyft perception dataset has a number of vehicles
that were not labeled. We plan to open source the errors we have found to
aid in the development of consistent labeling for the Lyft dataset.

\subsection{\sn can Find Missing Observations}
We additionally searched for missing observations in human-proposed tracks as a
case study.
To find missing human-proposed labels within tracks, we applied the following
AOF. We set the probability of an observation in a bundle with a human proposal
to 0. We set the probability of any track without a human proposal to 0. For
\sn, we then ranked the bundles by likelihood. For the ad-hoc MA baseline, we
random ordered bundles.

Within the datasets, we were only able to find a single example of such a
missing observation. For this example, \sn ranked the missing observation at the
top. We show the missing observation in Figure~\ref{fig:missing_obs} and
examples of low probability missing observations in
Figure~\ref{fig:mis_specification}. The feature distributions correctly identify
consistent predictions within tracks and correctly downweights inconsistent
predictions.

\begin{figure*}[t!]
    \centering
    \begin{subfigure}{0.33\textwidth}
        \includegraphics[width=\textwidth]{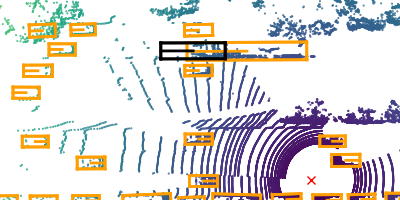}
        \caption{$t=0$}
    \end{subfigure}
    \begin{subfigure}{0.33\textwidth}
        \includegraphics[width=\textwidth]{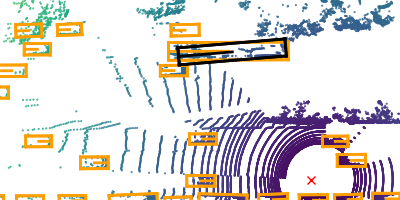}
        \caption{$t=1$}
    \end{subfigure}
    \begin{subfigure}{0.33\textwidth}
        \includegraphics[width=\textwidth]{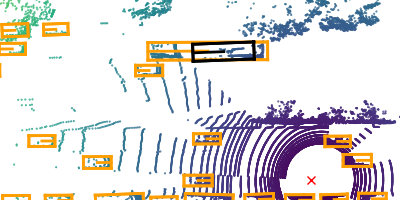}
        \caption{$t=2$}
    \end{subfigure}
    \caption{Example of a model error (in black) in the Lyft dataset not found
    by ad-hoc model assertions. We show ground-truth boxes from human labels in
    orange for reference. The erroneous prediction overlaps across frames, but
    is not consistent. \sn can find such errors as they produce unlikely values
    under learned feature distributions.}
    \label{fig:model_error}
\end{figure*}





\subsection{\sn can Find Novel ML Prediction Errors}
We further investigated whether or not \sn can find errors in LIDAR model
predictions. For this use-case, we did not assume access to human-proposed
labels.

\minihead{Experimental setup}
Unlike for finding errors in human-proposed labels, ad-hoc MAs can achieve high
precision when searching for errors in ML model predictions. As such, we
deployed three ad-hoc MAs as used in \citet{kang2020model} (\texttt{appear},
\texttt{flicker}, and \texttt{multibox}). Briefly, the \texttt{appear} assertion
asserts that an observation should have observations in nearby timestamps, the
\texttt{flicker} assertion asserts than an observation should not appear and
disappear rapidly, and the \texttt{multibox} assertion asserts that 3 boxes
should not overlap. These assertions can be reproduced in \sn with hand-tuned
features.

In addition to ad-hoc MAs, we additionally compared to uncertainty sampling, in
which we sampled predictions around a confidence threshold.

We then deployed \sn to find errors in ML model prediction after \emph{excluding
the errors found by these ad-hoc MAs}. We searched for both localization and
classification errors.  For \sn, we deployed the same features as above with the
exception of distance and model only. We additionally deployed a track feature
over the total number of observations. We measure the precision of the top 10
potential errors over 5 scenes in the Lyft dataset.

\minihead{Results and discussion}
Across these scenes, \sn achieves a precision at 10 of 82\% while uncertainty
sampling achieved a precision of 42\%. We note that errors we found with \sn
were not found by ad-hoc MAs. Many of these errors have tracks without missing
time stamps (so will not trigger the \texttt{flicker} assertion) and are longer
than two observations (so will not trigger the \texttt{appear} assertion). We
show an example of such a track in Figure~\ref{fig:model_error}, in which the
predictions overlap across frames, but in an unlikely way.

Furthermore, in contrast to uncertainty sampling, \sn uncovers errors with high
model confidence. \sn discovered errors in ML model predictions with confidences
as high as 95\%, which uncertainty sampling would miss.

\section{Related Work}

\minihead{Data cleaning}
Data quality is of paramount concern, especially in ML pipelines. Work in the data systems community has focused on cleaning tabular data~\cite{chu2016data, rahm2000data}. This work focuses largely on detecting errors via constraints~\cite{bertossi2006consistent, beskales2010sampling, bohannon2005cost} and more recently machine learning~\cite{krishnan2016activeclean, rekatsinas2017holoclean, heidari2019holodetect}. Unfortunately, these techniques do not directly apply to the labels in many ML pipelines, thus necessitating the need for new abstractions and systems for ML data.

\minihead{Enriching data and weak supervision}
Other work in the data systems community aims to clean or enrich data, often with the goal of training ML models. For example, HoloClean automatically aggregates noisy cleaning rules in a statistical fashion \cite{rekatsinas2017holoclean, heidari2019holodetect}. In this work, we focus on high quality training data in mission-critical settings as opposed to noisy cleaning rules.

Other work uses weak labels or organizational resources to train models with little labeled data. For example, users can specify noisy labeling functions that are aggregated to train models \cite{ratner2020snorkel}. Other work uses organizational resources, e.g., embeddings and knowledge bases, to adapt models to new modalities \cite{suri2020leveraging}. In mission-critical pipelines, these methods still need to be validated with a high-quality dataset, so require labels~\cite{ratner2020snorkel}. We leverage different organizational resources (existing labels and models) to find errors in labels, as opposed to training new models with fewer labels.

\minihead{ML testing}
%
According to a recent survey \cite{zhang2020machine}, existing work in ML testing focuses on pipelines where schemas have meaningful information, such as categorical or numeric data \cite{hynes2017data, polyzotis2019data, polyzotis2017data}. While important for deployments with schemas, they do not apply to the settings we consider. Other work considers statistical measures of accuracy \cite{amershi2019software}, fuzzing for numeric errors \cite{odena2019tensorfuzz}, worst case perturbations \cite{xiang2018verification}, data linting \cite{hynes2017data}, and other techniques \cite{pham2019cradle}. These approaches are complementary to \sn.

In this work, we primarily focus on finding errors in complex perception training data and model errors. To our knowledge, model assertions are closest line of work; see Section~\ref{sec:background} for an extended discussion of MAs \cite{kang2020model}. Briefly, users must manually specify MAs and severity scores, which can be challenging in practice and miss important classes of errors.


\minihead{Factor graphs}
The closest analog of \dsl in the Bayesian setting are factor graphs
\cite{dellaert2017factor, kschischang2001factor} that are widely used in robot
perception. Factor graphs are a probabilistic tool to encode and factorize the
joint distribution of random variables as a product of locally, conditionally
independent functions. Modern mapping \cite{mur2015orb}, tracking
\cite{poschmann2020factor}, and localization \cite{dellaert2017factor} in robot
perception use factor graphs to incorporate spatiotemporal, and multi-modal
measurements into a probabilistic framework for Bayesian interpretation. We
formulate a similar graphical framework to jointly reason over domain-specific
feature distributions and application objective functions, composing together to
form MA graphs. However, unlike robot perception applications that incorporate
raw sensor measurements as individual factors, we incorporate both model
predictions and object priors as factors in \sn in order to acausally reason
over the individual model prediction measurements. 

\section{Discussion and Future Work}

We believe that \sn is an exciting first step towards data management for
complex, unstructured ML data. However, while effective at its particular task,
many questions remain.

For example, \sn was tailored for autonomous vehicle sensor data, but it may
also be applicable to other domains with temporal aspects, such as audio or
time series data. We view ML data management in a variety of domains, both for
unstructured and structured data, to be an exciting area of future research.

\sn also assumes independence between features and that features are not
misspecified. While these assumptions produce reasonable results, more
sophisticated graphical models and learning may help in ML data management.

\section{Conclusion}

To address the problem of finding errors in labels and ML models, we propose
Learned Observation Assertions (\dsl) and implement \sn. \dsl allows users to
specify data-driven feature distributions to indicate which data points are
potentially erroneous. Our prototype implementation of \dsl, \sn, leverages
existing organizational resources (trained ML models and existing labeled data)
to find labeling errors. We show that \sn can find errors in human labels up to
2$\times$ more effectively than prior research work.


\bibliographystyle{ACM-Reference-Format}
\bibliography{paper}


\end{document}